\documentclass[%
reprint,
superscriptaddress,
amsmath,amssymb,
aps,
]{revtex4-1}

\usepackage{color}
\usepackage{graphicx}
\usepackage{dcolumn}
\usepackage{bm}

\begin {document}

\title{Universal Fluctuations of Single-Particle Diffusivity in Quenched Environment
}

\author{Takuma Akimoto}
\email{akimoto@keio.jp}
\affiliation{%
  Graduate School of Science and Technology, Keio University, Yokohama, 223-8522, Japan
}%

\author{Eli Barkai}
\affiliation{%
  Department of Physics, Bar Ilan University, Ramat-Gan 52900, Israel
}%

\author{Keiji Saito}
\affiliation{%
  Department of Physics, Keio University, Yokohama, 223-8522, Japan
}%


\date{\today}

\begin{abstract}
Local diffusion coefficients in disordered materials such as living cells are highly heterogeneous. Quenched disorder 
is utilized substantially to study such complex systems, whereas its analytical treatment is difficult to handle. We consider 
finite systems with quenched disorder 
in order to investigate the effects of sample disorder fluctuations and confinement on single-particle diffusivity. 
While the system is ergodic in a single disorder realization,   
the time-averaged mean squared displacement depends on the disorder, i.e., the system is ergodic but non-self-averaging. We find that 
the inverse L\'evy distribution is a universal distribution for diffusivity in the sense that it can be applied for 
arbitrary dimensions. Quantifying the degree of the non-self-averaging effect, we  show that fluctuations of single-particle diffusivity
 far exceed the corresponding annealed theory and also find confinement effects. 
The relevance for experimental situations is also discussed.
\end{abstract}

\maketitle


{\it Introduction}.---Anomalous diffusion, where the mean square displacement (MSD) does not 
depend linearly on time, unlike Brownian motion,  has been extensively observed in complex systems such as disordered 
materials \cite{Scher1975, bouchaud90} and living cells \cite{Golding2006, Weigel2011, Manzo2015}. 
One of the origins of anomalous diffusion is ascribed to a quenched random environment with highly heterogeneous  
local diffusion coefficients. Such heterogeneous environments 
play a crucial role in the fluctuations of diffusivity observed in one-dimensional diffusion of proteins on DNA \cite{Graneli2006, Wang2006} 
and diffusion in living cells \cite{Manzo2015}. In fact, recent experiments have clearly demonstrated that diffusivity maps for cells 
are heterogeneous \cite{Kuhn2011, Masson2014}.

In single-particle-tracking experiments, the trajectory ${\bm r}(t)$ of a tracer in a medium is recorded. 
One of the most common tools to quantify the diffusivity is the time-averaged MSD:
\begin{equation}
 \label{tamsd_definition}
  \overline{\delta^{2}(\Delta;t)} \equiv
  \frac{1}{t - \Delta} \int_{0}^{t - \Delta} dt' \,
  [{\bm r}(t' + \Delta) - {\bm r}(t')]^{2} .
\end{equation}
For Brownian motion in a homogeneous medium, the time-averaged MSD converges to the ensemble average MSD
in the limit of long measurement time $t$, 
and the diffusion is then normal.
In strongly disordered systems, this equivalence 
can be broken, which is usually observed together with the onset 
of anomalous diffusion \cite{Metzler2014, Hofling2013, Meroz2015}.

 Anomalous diffusion in a quenched environment is sometimes discussed by replacing the quenched disorder by an annealed one, that is, 
  the continuous-time random walk (CTRW) approximation is employed \cite{Scher1975, bouchaud90}. This framework can capture many physical 
 features of anomalous diffusion especially in infinite systems far from equilibrium \cite{Scher1975, Wong2004}. 
 However, when we look at finite size disordered systems such as proteins on DNA or in living cells, 
 it is not clear whether the annealed picture can accurately describe the underlying diffusion processes \cite{Graneli2006, Wang2006, Manzo2015}. 
 Therefore, it is desired to clarify properties of single-particle diffusion that are inherent in quenched environment. 

To consider single-particle tracking in quenched disordered systems, one must take into account  four averaging procedures. 
The first is an  average of an observable over time, e.g. Eq.~(\ref{tamsd_definition}).
 The second is an  average over thermal paths namely an average over repeated experiments in the same realization of disorder. 
The third one is an average with respect to the quenched environment, i.e., the disorder average. 
The fourth procedure is with respect to the initial condition, and here one usually considers two choices:
 a particle initially in equilibrium with its environment,
or a particle initially injected into the system at random location. 
For finite systems in equilibrium, the initial condition is given by Boltzmann statistics. 
When the system is ergodic,  the long-time average and the equilibrium ensemble average (i.e., thermal-path average 
with an equilibrium initial condition) are equivalent. However, even when the system is ergodic,
 the equilibrium ensemble average may depend strongly on the disorder, which means 
that the sample-to-sample fluctuations remain large even when the system size is increased 
 (non-self-averaging property) \cite{bouchaud90, Aharony1996, Wiseman1998}. 
 An important question arises here: 
 Is there crucial discrepancy of the fluctuations of diffusivity in between quenched system and the corresponding annealed system? 
 To answer this question, we quantify the degree of non-self-averaging property. 

In this Letter, we consider the quenched trap model (QTM) \cite{bouchaud90}, and derive several rigorous and 
universal properties that characterize the anomalous diffusion in the quenched disorder. 
We show that fluctuations in single-particle tracking  
in quenched environment, among different realizations of the disorder, far exceed 
the corresponding fluctuations found for the annealed CTRW case. Thus, 
against common belief,  
the annealed model like CTRW does not capture main ingredients of anomalous diffusion in the quenched environment. 
We show that the exact statistics of fluctuations of diffusivity is universal, because it is valid for any dimension. 
Confinement effects are also demonstrated.  
These will provide a basis to consider anomalous diffusion of single particles in finite systems with quenched disorder.

{\it Model}.---We consider a random walk on a quenched random energy landscape 
on a finite $d$-dimensional hypercubic lattice 
\cite{bouchaud90}. Quenched disorder means that when realizing the random energy landscape it does not change with time.   
The lattice constant is set to unity and the number of  lattice sites with different energies is finite, e.g., ${\bm r}_k=1,2, \cdots, L$ 
($k=1, \cdots, d$).
At each lattice point, the depth $E>0$ of an energy trap is randomly assigned. The depths are 
independent identically distributed random variables with an exponential distribution,
$\rho(E) = T^{-1}_g \exp(-E/T_g)$. A particle can escape from a trap and jump to one of the nearest neighbors. 
The mean trapping time $\tau_{\bm r}$ at site ${\bm r}$ follows Arrhenius law, i.e., $\tau_{\bm r} = \tau_0 \exp(E_{\bm r}/T)$, 
where $E_{\bm r}$ is the depth of the energy at site ${\bm r}$, 
$T$  the temperature, and $\tau_0$ a typical time. 
It is easy to show that the probability density function (PDF), $\psi_\alpha(\tau)$, of trapping times follows
\begin{equation}
\int_\tau^\infty d\tau' \psi_\alpha (\tau') = \left(\frac{\tau_0}{\tau} \right)^{-\alpha}\quad (\tau \geq \tau_0)
\label{power-law-pdf}
\end{equation}
with $\alpha \equiv T/T_g$ \cite{Bardou2002}. 
Thus, the mean trapping time diverges for $\alpha \leq 1$, which leads to anomalous behaviors 
\cite{Machta1985, bouchaud90, Monthus1996, Burov2007, *Burov2011, Miyaguchi2011, Massignan2014, Miyaguchi2015, Luo2015}. 
We note that the sample mean trapping time $\mu = \sum_{\bm r} \tau_{\bm r} /L^d$
for a fixed disorder never diverges when $L$ is finite. 
Thus, the process can reach an equilibrium state and present ergodic behavior with the aid of 
the finite characteristic time scale of the system.

CTRW is an annealed model which mimics certain aspects of dynamics of the QTM. In CTRW the particle jumps between nearest neighbors with waiting times
drawn from Eq.~(\ref{power-law-pdf}), and  the waiting time distributions at all lattice points are identical. In that sense the 
system is homogeneous. 

For the QTM with a finite lattice size $L$, we can consider Boltzmann statistics (equilibrium statistical physics).  
Let $P_{\bm r}$ be the probability of finding a particle at site ${\bm r}$. 
Except for the boundary, the master equation for the $i$th single disorder realization $\tau_{\bm r}^{(i)}$ 
is given by
\begin{equation}
\frac{dP_{\bm r}}{dt} = \frac{1}{2d} \sum_{\bm r'}\frac{P_{\bm r'}}{\tau_{\bm r'}^{(i)}}  - \frac{P_{\bm r}}{\tau_{\bm r}^{(i)}},
\end{equation}
where the sum is  over the nearest neighbor sites. 
We consider two boundary conditions:  periodic and reflecting. 
In the periodic boundary condition, the energies in the random energy landscape are periodically arranged. 
In the reflecting boundary condition, a particle will return to the original position when it hits  the boundary. 
In both cases, one obtains the equilibrium state 
\begin{equation}
P_{\bm r}^{\rm eq} = \frac{\tau_{\bm r}^{(i)}}{L^d \mu_i},
\label{eq_state}
\end{equation}
where $\mu_i$ is the sample mean trapping time 
for the $i$th single disorder realization,  i.e., $\mu_i = \sum_{\bm r}\tau_{\bm r}^{(i)} /L^d$.  
In what follows, we consider the equilibrium distribution (\ref{eq_state}) as an initial distribution. 

{\it Universal distribution of diffusion coefficient}.---Here, we consider the periodic boundary condition. The
MSD for the $i$th disorder realization increases as $\langle \{{\bm r}(t) - {\bm r}(0)\}^2 \rangle_{\rm eq} = \langle N_t \rangle_{\rm eq}$,
where $\langle N_t \rangle_{\rm eq}$ is the mean number of jumps until time $t$ and $\langle \cdot \rangle_{\rm eq}$ 
implies the equilibrium ensemble average.
At equilibrium, $\langle N_t \rangle_{\rm eq}$ is given by
\begin{equation}
\langle N_t \rangle_{\rm eq} = \frac{t}{\mu_i}
\label{renewal_function}
\end{equation}
for a specific disorder realization. This result is exact for any $t>0$. 
We note that this average is taken over equilibrium initial conditions and thermal histories but 
not over disorder.  

\begin{figure}
\includegraphics[width=.9\linewidth, angle=0]{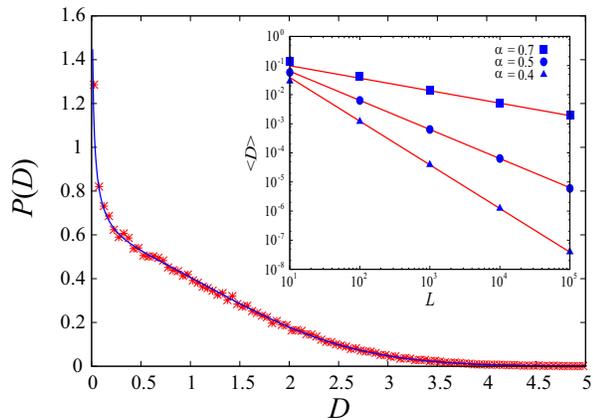}
\caption{Distribution of the diffusion coefficients for different disorder realizations ($T=1$ and $T_g=1.5$). 
The crosses are the results of the numerical simulation  ($d=1$ and $L=10^4$).  In the numerical simulation, we calculated the distribution 
of $D=1/\mu^i$ for different disorder realizations (see \cite{SM} for finite time simulations).
The mean of the PDF is set to  unity. The solid line is the inverse 
L\'evy distribution, Eq. (\ref{PDF_inverse_Levy}). The inset shows
the disorder average of diffusion coefficients as a function of the system size $L$ for several $\alpha=T/T_g$ ($T_g=1$). 
Here, the symbols are the results of numerical simulations and   
the solid lines are the theoretical curves, Eq.~(\ref{D_dis_ave}).
 }
\label{dist_diffusion}
\end{figure}

Because the disorder $\tau_{\bm r}^{(i)}$ is periodically arranged, the MSD grows as 
 $\langle \{{\bm r}(t) - {\bm r}(0)\}^2 \rangle_{\rm eq}=t/\mu_i$. We define
the diffusion coefficient for a single disorder realization $i$ as $D_i = 1/\mu_i$. 
By the law of large numbers, for $\alpha > 1$, 
we have
$(\tau_0^{(i)} + \cdots + \tau_{L^d-1}^{(i)})/L^d \rightarrow  \langle \tau \rangle$  ($L\to \infty$),
where $\tau_k^{(i)}$ is a trapping time at site $k = \sum_{l=1}^{d} L^{l-1}({\bm r}_l -1)$ and 
$\langle \tau \rangle \equiv \int_0^\infty  \tau \psi(\tau) d\tau$. 
Because $\langle \tau \rangle$ is determined uniquely by $\alpha$, the diffusion coefficient does not depend on the disorder sample.  
This is a consequence of the self-averaging property. 
On the other hand, 
because 
the law of large numbers breaks down for $\alpha \leq 1$,  
 the PDF of the sum of $\tau_{k}^{(i)}$ 
follows the one-sided L\'evy distribution \cite{Feller1971}:
\begin{equation}
\frac{\tau_0^{(i)} + \cdots + \tau_{L^d-1}^{(i)}}{(L^d)^{1/\alpha}} \Rightarrow X_\alpha\quad (L\to \infty),
\end{equation}
where $X_\alpha$ is a random variable following the one-sided L\'evy distribution of index $\alpha$.
The PDF of $X_\alpha$ denoted by 
$l_\alpha(x)$ with $x>0$ is given by \cite{Feller1971}
\begin{equation}
l_\alpha(x) = -\frac{1}{\pi x} \sum_{k=1}^\infty \frac{\Gamma(k\alpha +1)}{k!} (-cx^{-\alpha})^k \sin (k\pi \alpha),
\label{levy_pdf}
\end{equation}
where $c=\Gamma (1-\alpha) \tau_0^\alpha$ is a scale parameter. 
Here, we define the inverse L\'evy distribution as the PDF of $X_\alpha^{-1}$:
\begin{equation}
g_\alpha(y) = -\frac{1}{\pi y} \sum_{k=1}^\infty \frac{\Gamma(k\alpha +1)}{k!} (-cy^{\alpha})^k \sin (k\pi \alpha).
\label{PDF_inverse_Levy}
\end{equation}
Because the diffusion coefficient is given by $D_i = L^{1-1/\alpha} X_\alpha^{-1}$,  the PDF of $D_i$ 
is described by the inverse L\'evy distribution and hence $D_i$ depends crucially on the sample of the disorder realization. 
The first and the second moments of the inverse L\'evy distributions are calculated in \cite{SM}.
As shown in Fig.~\ref{dist_diffusion}, our rigorous  result for the distribution of the diffusion coefficients is in good agreement with 
the numerical simulations. 
Surprisingly, the inverse L\'evy distribution is a universal distribution of the diffusion coefficient in the sense that it is exact for any dimension. 
Using the first moment of the inverse L\'evy distribution, we obtain the exact expression of 
the disorder average of the diffusion coefficient:
\begin{equation}
\langle D \rangle_{\rm dis} = \frac{L^{1-1/\alpha}\Gamma(\alpha^{-1})}{\alpha \tau_0 \Gamma(1-\alpha)^{1/\alpha}},
\label{D_dis_ave}
\end{equation}
where $\langle \cdot \rangle_{\rm dis}$ means the disorder average. 
This result perfectly matches the simulation presented in Fig.~\ref{dist_diffusion}. 
We note that the disorder average of the diffusion coefficient depends on the size of the system and it becomes zero 
as the system size $L$ goes to infinity.

{\it Ergodicity versus self averaging}.---To investigate the ergodic properties of
the disordered system, we consider the ergodicity breaking (EB) parameter \cite{He2008} defined by 
\begin{equation}
{\rm EB}(t;\Delta)\equiv \frac{\langle \{\overline{\delta^{2}(\Delta;t)}\}^2 \rangle_{\rm eq}  - \langle \overline{\delta^{2}(\Delta;t)} \rangle_{\rm eq}^2}
{\langle \overline{\delta^{2}(\Delta;t)} \rangle_{\rm eq}^2}.
\label{EB}
\end{equation} 
If the EB parameter goes to zero, the time-averaged MSD for a single disorder realization 
converges to the equilibrium ensemble average, that is, the process is ergodic:
$\overline{\delta^{2}(\Delta;t)} \to \langle {\bm r}(\Delta)^2 \rangle_{\rm eq}$ for $\Delta >0$ ($t\to\infty$).
In CTRW, the EB parameter is not zero even when $t$ goes to infinity \cite{He2008, Miyaguchi2013}. 
For $1 \ll \Delta \ll t$, 
the EB parameter for a single disorder realization decays as
\begin{equation}
{\rm EB}(t;\Delta) \sim \frac{4\Delta}{3dt} \quad (t \to \infty ~{\rm and}~\Delta \gg 1),
\label{EB_qtm}
\end{equation} 
which means that the system is ergodic (see \cite{SM}). 
This statement becomes invalid for infinite system ($L=\infty$) because 
there is no equilibrium state in that case.

\begin{figure}
\includegraphics[width=.9\linewidth, angle=0]{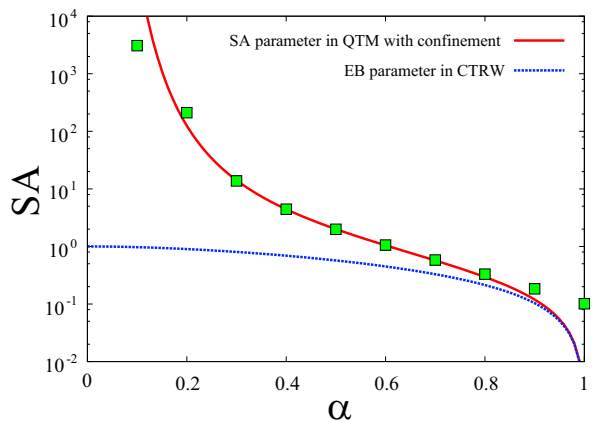}
\caption{Self-averaging parameter as a function of $\alpha$. The symbols are the result of a numerical 
simulation ($d=1$ and $L=10^4$). The solid line is the theory, Eq. (\ref{SA_theory}). The dotted line is the EB parameter 
in CTRW \cite{He2008}. 
 }
\label{SA_L=1000}
\end{figure}

Next, we propose another quantity characterizing the self-averaging property, coined the self-averaging (SA) parameter, defined 
by
\begin{equation}
{\rm SA}(t,L;\mathcal{O})\equiv \frac{\langle \overline{\mathcal{O}(t)}^2 \rangle_{\rm dis}  - \langle \overline{\mathcal{O}(t)} \rangle_{\rm dis}^2}
{\langle \overline{\mathcal{O}(t)} \rangle_{\rm dis}^2 },
\label{SA}
\end{equation} 
where $\overline{\mathcal{O}(t)}$ is a time-averaged observable, i.e., $\overline{\mathcal{O}(t)}\equiv \int_0^t dt' \mathcal{O}(t')/t$. 
If the SA parameter becomes zero for the limits $t\to\infty$ 
and $L\to \infty$, the system is called self-averaging because the fluctuations of the time-averaged observable due to different disorder realizations 
disappear when the systems become large. 
Note that self-averaging property in finite systems can be characterized by the asymptotic limit of $L$ when 
the limit $L\to \infty$ is taken after the limit $t\to\infty$. 
Because the system is ergodic for finite $L$,  the SA parameter for time-averaged MSD becomes 
\begin{align}
 {\rm SA}(t,L;\delta{\bm r}^2_\Delta) 
&= \frac{\langle \overline{\delta^{2}(\Delta;t)}^2 \rangle_{\rm dis}  - \langle \overline{\delta^{2}(\Delta;t)}\rangle_{\rm dis}^2}
{\langle \overline{\delta^{2}(\Delta;t)}\rangle_{\rm dis}^2 } \nonumber\\
&\to \frac{\langle 1/\mu_i^2 \rangle_{\rm dis}  - \langle 1/\mu_i \rangle_{\rm dis}^2}{\langle 1/\mu_i \rangle_{\rm dis}^2 } 
\quad (t\to\infty),
\end{align}
where $\delta {\bm r}_\Delta \equiv {\bm r}(t+\Delta) - {\bm r}(t)$. 
Using the first and second moment of $1/\mu_i$ obtained in \cite{SM}, we have 
\begin{equation}
\lim_{L\to \infty} \lim_{t\to \infty} {\rm SA}(t,L;\delta{\bm r}^2_\Delta) 
=
\left\{
\begin{array}{ll}
0 &(\alpha >1)\\
\\
\frac{\alpha \Gamma(\frac{2}{\alpha}) }{\Gamma(\frac{1}{\alpha})^2} -1\quad &(\alpha \leq 1).
\end{array}
\right.
\label{SA_theory}
\end{equation}
It follows that the system is not self-averaging for $\alpha<1$, whereas it is ergodic when $L<\infty$. 
The results obtained so far show striking differences if compared with CTRW. 
 In CTRW one finds ergodicity breaking 
\cite{Bouchaud1992, Lubelski2008, He2008, Miyaguchi2013} while
 so far we have found non-self averaging. Importantly the fluctuations in the quenched model are exponentially
 larger than the annealed model. This is quantified by a very large SA parameter, if compared with the EB parameter of CTRW (see Fig.~\ref{SA_L=1000}). 
 Furthermore, the distribution of the diffusion constant is not bounded at $D=0$
(see Fig.~\ref{dist_diffusion}), which implies a heavy statistical  weight
for very slow particles. Because this effect is not found for the annealed model, quenched models lead to surprisingly large fluctuations. 
Finally, in CTRW theory, the diffusion coefficient depends on the measurement time, that is, a phenomenon called aging 
\cite{Metzler2014, He2008, Lubelski2008, Schulz2013}. 
On the other hand, for finite size system with quenched disorder, the system size controls the long time statistics of the 
diffusion coefficient, e.g. Eq.~(\ref{D_dis_ave}).

{\it Effect of confinement}.---For the reflecting boundary condition, the MSD converges to a constant as time goes to infinity due to the confinement, 
while it increases as $\langle \{{\bm r}(t) - {\bm r}(0)\}^2 \rangle_{\rm eq} \sim \langle N_t \rangle_{\rm eq}$ for short $t$. 
  Because the system is in equilibrium, 
 the constant is given by $\sigma_i^2 = 2(\langle {\bm r}^2 \rangle_{\rm eq} - \langle {\bm r} \rangle_{\rm eq}^2)$, which is a non self-averaging variable 
 when $T\leq T_g$.
If we define the crossover time $t_c$ from the diffusive to plateau regime as $\langle N_{t_c} \rangle = \sigma_i^2$,  we 
have $t_c=\mu_i \sigma_i^2$. 
Since the MSD depends on the disorder of random energy landscape, the crossover time is also fluctuating.

Since the system is ergodic for a single disorder realization, time average converges to the equilibrium ensemble average:
$\overline{ {\bm r}(t)} \equiv \int_{0}^{t} {\bm r}(t') dt' /t  \to \langle {\bm r} \rangle_{\rm eq}$ and 
$\overline{ {\bm r}^2(t)} \equiv \int_{0}^{t} {\bm r}(t')^2 dt' /t  \to \langle {\bm r}^2 \rangle_{\rm eq}$ as $t\to \infty$. 
When the value of the observable is determined by the site ${\bm r}$, i.e., $\mathcal{O}_{\bm r}$, 
the time-averaged observables can be represented by the equilibrium probability:
$
\overline{\mathcal{O}} = \sum_{\bm r} \mathcal{O}_{\bm r} P^{\rm eq}_{\bm r} = \frac{ \sum_{\bm r} \mathcal{O}_{\bm r} \tau_{\bm r}}{\sum_{\bm r} \tau_{\bm r}}.
$ 
We note that these time averages depend strongly on the disorder for $\alpha<1$. 
Using methods similar to those presented in \cite{Rebenshtok2007, *Rebenshtok2008}, 
we show in \cite{SM} that the SA parameter for position is 
\begin{align}
\lim_{L\to \infty} \lim_{t\to \infty} {\rm SA}(t,L; {\bm r}) 
 &= \lim_{L\to \infty} \frac{\langle \langle {\bm r}\rangle_{\rm eq}^2 \rangle_{\rm dis}  - 
 \langle \langle {\bm r}\rangle_{\rm eq}\rangle_{\rm dis}^2}{\langle \langle {\bm r}\rangle_{\rm eq} \rangle_{\rm dis}^2 } \nonumber\\
 &=
\left\{
\begin{array}{ll}
0 &(\alpha >1)\\
\\
\frac{1-\alpha}{3}\quad &(\alpha \leq 1).
\end{array}
\right.
\label{SA_position}
\end{align}
Thus, the non-self-averaging behavior of the position under confinement appears for $\alpha < 1$. 
Unlike the SA parameter for the time-averaged MSD,  the SA parameter does not blow up when $\alpha \to 0$. 
This is likely because we are dealing here with an equilibrium observable which is time-independent. 

{\it Discussion}.---We analytically showed ergodicity and non-self-averaging properties in $d$-dimensional QTM in a finite system. 
The transition from self-averaging to non-self-averaging occurs at $\alpha =1$, i.e., $T=T_g$. 
Non-self averaging is a consequence of the breakdown of the central limit theorem for the waiting times at sites. 
As a result, the non-self-averaging effects lead to universal fluctuations of diffusivity, that is, the PDF of the 
diffusion coefficient follows the inverse L\'evy distribution in arbitrary dimension, which is different from the annealed model (CTRW).
The inverse L\'evy distribution stems from the L\'evy distribution, which is a universal distribution for the sum 
of waiting times. Therefore, it will be found in other models beyond the QTM like the random comb model and 
 the results are truly universal. 
We also quantified the degree of the non-self-averaging property by the SA parameter and found a large difference from 
that in the annealed model  (see Fig.~\ref{SA_L=1000}). Note that the same averaging procedure is used to calculate 
the EB parameter in CTRW, and hence it is significant to compare the SA parameter in QTM with the EB parameter.
The  quenched and annealed systems exhibit similar type of randomness of diffusion constants only for infinite systems 
and in dimension $d>2$. 

\begin{figure}
\includegraphics[width=1.\linewidth, angle=0]{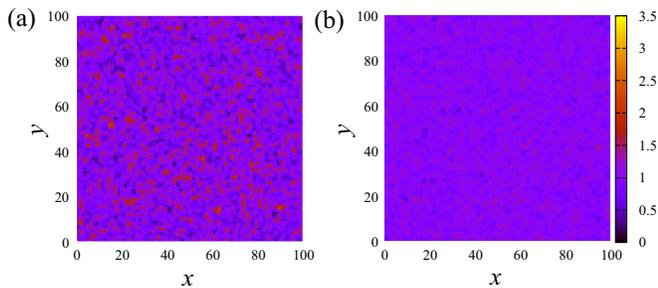}
\caption{Two-dimensional diffusivity maps  (a) $\alpha=0.8$ and (b) $\alpha=1.5$ ($L=900$). 
Diffusivity at the site $(i,j)$ is represented by the time-averaged MSD with $\Delta=1$ divided by the mean, 
$t=10^3$, and initial points $( 9(i-1) + 5, 9(j-1) +5)$ for $i,j = 1, \cdots, 100$. We use a coarse-graining of sites in the figure, i.e., one site in the figure 
contains $9\times 9$ sites.
 }
\label{diffusivity_map}
\end{figure}

There are many biological experiments  described by quenched environment 
with heterogeneous local diffusivity \cite{Graneli2006, Wang2006, Kuhn2011}. 
In experiments so far, one uses diffusion maps to characterize the heterogeneity of the system. 
Figure~\ref{diffusivity_map} presents 
the local diffusivity defined as the time-averaged MSD with a fixed $\Delta$ divided by the mean, where the ensemble of the mean 
is over a uniform initial ensemble in a single disorder realization. 
The diffusivity map becomes highly heterogeneous when $\alpha$ is smaller than one. This heterogeneity results from the random energy 
landscape because the local diffusivity is correlated with the energy (deep energy trap implies slow diffusivity). 
While the diffusivity map in CTRW is also heterogeneous, similar to that in the corresponding QTM, 
 it is not reproducible because of the annealed picture and hence it is meaningless. 
Therefore, the reproducible property of the diffusivity map, which is absent in the annealed picture, play an important role in capturing 
the heterogeneity in the quenched environment.

Diffusion in quenched systems exhibits an effect known as population splitting \cite{Schulz2013}. 
As shown in Fig.~\ref{dist_diffusion}, PDF of the diffusion coefficients becomes unbounded at $D=0$, which cannot be observed 
in the annealed version (CTRW). We confirm numerically a similar behavior in a finite system and finite measurement times
(see Fig.~S4 in the Supplementary Material \cite{SM}). 
 Namely, particles split into immobile and mobile particles in a finite system with quenched disorder.

In 2008 it was claimed that nonergodicity (found in CTRW) mimics inhomogeneity, where the time-averaged MSDs for 
different realizations exhibit large fluctuations \cite{Lubelski2008, He2008}. In this Letter, we have obtained universal 
distributions to describe the fluctuations of the inhomogeneous system.  We have shown that starting from a thermal state and for a finite though 
large  system the fluctuations stemming from inhomogeneity far exceed those obtained from the simpler annealed model. 
Thus, the annealed approach hides rich physical behaviors that are now quantified.  


%



\end{document}